\documentclass{article}
\usepackage{waspaa23,amsmath,graphicx,url,times}
\usepackage{color}
\usepackage{booktabs}
\usepackage{caption}
\usepackage{ifthen}
\title{Bridging High-Quality Audio and Video via Language\\ for Sound Effects Retrieval from Visual Queries}

\name{Julia Wilkins$^{1, 2}$
      Justin Salamon$^{2}$
      Magdalena Fuentes$^{1}$
      Juan Pablo Bello$^{1}$
      Oriol Nieto$^{2}$}
\address{$^1$ Music and Audio Research Laboratory, New York University, NY, USA\\
         $^2$ Adobe Research, San Francisco, CA, USA\\
         \ninept{\url{jw3596@nyu.edu}}\\
          \ninept\textcolor{magenta}
         {\url{https://juliawilkins.github.io/sound-effects-retrieval-from-video/}}
}

\begin{document}
\pdfoutput=1
\ninept

\twocolumn[{%
\renewcommand\twocolumn[1][]{#1}%
\maketitle
\begin{center}
    \centering
  \includegraphics[width=\linewidth, trim={0 1 0cm 2cm}]{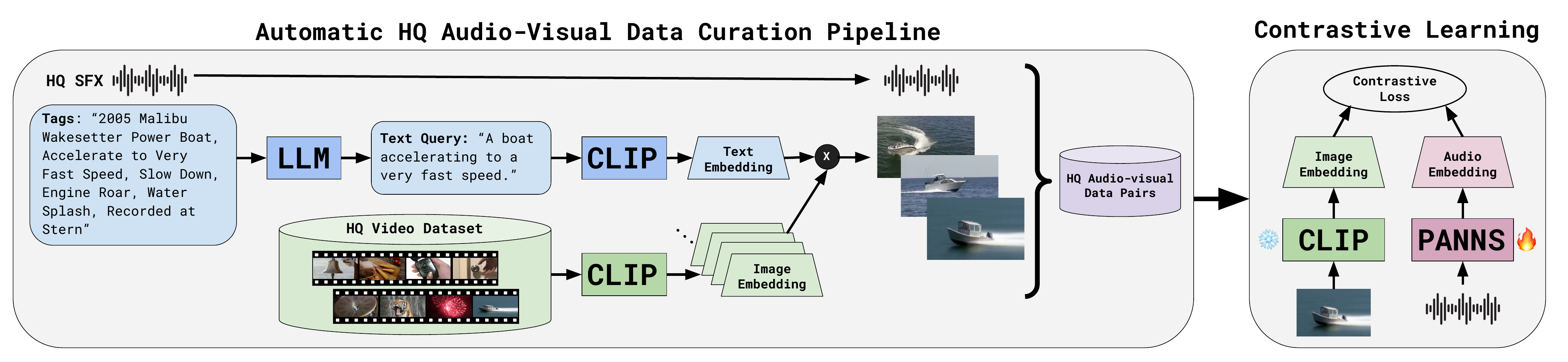}
  \captionsetup{type=figure}
  \captionof{figure}{
    Our automatic audio-visual data curation pipeline (left). The new data are used to learn multimodal audio-visual embeddings (right).
  }
  \label{fig:block_diag}

\end{center}%
}]
\begin{sloppy}

\vspace{-1cm}

\begin{abstract}
Finding the right sound effects (SFX) to match moments in a video is a difficult and time-consuming task, and relies heavily on the quality and completeness of text metadata. Retrieving high-quality (HQ) SFX using a video frame directly as the query is an attractive alternative, removing the reliance on text metadata and providing a low barrier to entry for non-experts. Due to the lack of HQ audio-visual training data, previous work on audio-visual retrieval relies on YouTube (``in-the-wild”) videos of varied quality for training, where the audio is often noisy and the video of amateur quality. As such it is unclear whether these systems would generalize to the task of matching HQ audio to production-quality video. To address this, we propose a multimodal framework for recommending HQ SFX given a video frame by (1) leveraging large language models and foundational vision-language models to bridge HQ audio and video to create audio-visual pairs, resulting in a highly scalable automatic audio-visual data curation pipeline; and (2) using pre-trained audio and visual encoders to train a contrastive learning-based retrieval system. We show that our system, trained using our automatic data curation pipeline, significantly outperforms baselines trained on in-the-wild data on the task of HQ SFX retrieval for video. Furthermore, while the baselines fail to generalize to this task, our system generalizes well from clean to in-the-wild data, outperforming the baselines on a dataset of YouTube videos despite only being trained on the HQ audio-visual pairs. A user study confirms that people prefer SFX retrieved by our system over the baseline 67\% of the time both for HQ and in-the-wild data. Finally, we present ablations to determine the impact of model and data pipeline design choices on downstream retrieval performance. Please visit our companion website to listen to and view our SFX retrieval results.
\end{abstract}

\begin{keywords}
Multimodal machine learning, cross-modal retrieval, audio-visual representation learning, data augmentation
\end{keywords}

\section{Introduction}
\label{sec:intro}

Video creation and editing commonly entails matching sound effects (SFX) to video to create an engaging multimedia experience. 
Yet finding the right SFX to pair with moments in a video is challenging and often time-consuming. Regardless of skill level, this process typically relies heavily on the quality and completeness of text metadata associated with an SFX library (i.e., for a text-based search), which can vary significantly by library. Further, it is common to use sounds from one object to represent entirely different objects (e.g., the sound of shoes on glass shards to represent walking through snow), an art known as ``Foley''~\cite{ament2021foley, de2013real}, meaning sonically relevant sounds may be missed if we rely on metadata for retrieval.

Multimodal machine learning methods have become popular across the visual, audio, and text domains~\cite{ngiam2011multimodal, ramachandram2017deep}, due to their ability to learn from the correspondence between signals in different modalities, often via contrastive learning~\cite{oord2019representation, chen2020simple, technologies9010002}, and the growing availability of large multimedia datasets~\cite{Ego4D2022CVPR, hebbar2023dataset, tian2018audio_AVE, chen2020vggsound, agostinelli2023musiclm}. Previous approaches for audio-visual retrieval \cite{wu2022wav2clip, guzhov2022audioclip, wu2023largescale, Zellers_2022_CVPR, krishnamurthy2021learning} rely on YouTube, aka ``in-the-wild”, videos \cite{chen2020vggsound, gemmeke2017audioset}, where the audio is often noisy and the video of amateur quality. These videos are not necessarily representative of the production-quality videos and associated audio that video creators often work with, so it is unclear whether systems trained on noisy video datasets can generalize to this type of high-quality (HQ) content. Datasets of HQ SFX moments in videos are lacking, though recent attempts have been made to address this by assuming sound event occurrences via video captions~\cite{hebbar2023dataset, nagrani2022learning} or even generation~\cite{10.1007/978-3-031-20074-8_25}. For music, unlike SFX, there are more HQ audio-visual data available (namely music video clips), which have been leveraged to train music retrieval systems for video \cite{pretet2021cross, Suris_2022_CVPR, mckee2023language}. To circumvent the need for HQ audio-visual pairs, Lin et al.~\cite{lin2021soundify} use CLIP~\cite{radford2021learning} to match embeddings of video frames against embeddings of text metadata associated  with audio SFX, to recommend SFX for a given video.

In this work, we address the lack of HQ audio-visual data for SFX via a novel automatic data curation pipeline, and use it to train an audio-visual SFX retrieval system that generalizes both to production-grade and in-the-wild media content. Our contributions can be summarized as follows:
(1) we propose an automatic data curation pipeline that bridges HQ audio and visual data via large language models and foundation vision-language models; (2) we show both quantitatively and qualitatively that by using this pipeline to train a contrastive audio-visual model, we are able to significantly outperform baselines on SFX retrieval given a visual input query; (3) we show that our model generalizes to noisy ``in-the-wild'' video despite being trained on the auto-curated HQ data; and (4) we present ablations to study the impact of key model and data pipeline design choices on downstream retrieval performance.

\vspace{-0.1cm}
\section{Method}

\label{sec:method}
 
\subsection{Automatic Audio-visual Data Curation Pipeline}
\label{sec:method_data}
To generate HQ audio-visual training data, we propose an automatic data curation pipeline that leverages language and vision-language models. First, we use text metadata associated with each SFX to generate text embeddings via CLIP \cite{radford2021learning}, and in parallel generate CLIP image embeddings of video frames. Next, we retrieve the most similar visual frames to each SFX-based text query via cosine similarity. This method is depicted in Figure 1 (left). Studies have shown that CLIP performs best at image retrieval when the input text is a clear natural language description \cite{zhou2022learning, yong2022prompt}. The text metadata in most HQ audio SFX libraries, including the one used for training in this work, does not fit this criterion. Rather, it is a list of tags (Figure \ref{fig:block_diag}, left) that may be over-specific or include audio-centric terms (e.g. ``left to right sound'') that are not helpful for retrieving relevant visual frames. To alleviate this, we propose to use a large language model (LLM) in the data curation pipeline to convert the tag data into natural language sentences before computing their CLIP embeddings. In this way, we obtain a text query that is both in the right format (natural language) and level of specificity for successful text-image matching using CLIP. We use Bloom \cite{scao2022bloom}, an open-source LLM with 176B parameters and leverage prompt analogies 
\cite{mckee2023language} to improve the quality of the output sentence. We found that prompt analogies, in addition to producing overall better quality sentences, allowed us to control their level of specificity and filter out information from over-detailed tags. Later in our experiments, we perform an ablation to capture the quantitative effect of using the LLM-transformed SFX tags for text-image matching as opposed to using the tags directly. 

Our approach takes inspiration from Lin et al. \cite{lin2021soundify}, with two key distinctions. First, we use an LLM to enhance the text metadata of SFX and optimize the image-text matching via CLIP. Most importantly, in \cite{lin2021soundify} they use this image-text relationship directly as the retrieval system. In our method, we use this as a way to generate training data to then train an audio-visual model such that we can retrieve SFX using visual queries \textit{without} the need for text. The entire pipeline is depicted in Figure \ref{fig:block_diag}.

The proposed data pipeline is fully automatic and thus highly scalable, constrained only by the size of the SFX and video datasets. Additionally, each SFX can be matched against the top $k$ relevant visual frames, serving as a form of data augmentation and producing a dataset of audio-visual pairs that is $k$ times the size of the SFX dataset. While our focus is on HQ SFX and video in this work, the pipeline is applicable to any combination of audio and visual data as long as the audio is associated with tag metadata.

\textbf{Datasets:} To create audio-visual training pairs using our pipeline, we draw data from high-quality SFX and video datasets. For SFX we use the ProSoundEffects (PSE) Core 3 Pro Plus \footnote{\url{https://www.prosoundeffects.com}.} bundle. 
It contains 150k professional-grade audio from over 80 SFX categories, along with metadata including category labels and notably a detailed list of metadata tags per audio file. Some files contain multiple SFX events separated by silence, in which case we segmented them into a total of 336k individual SFX.
For video, we use a large collection of over 1 million HQ stock videos\footnote{\url{https://stock.adobe.com/video.}} that are mostly 10 seconds or less in duration, and have associated text metadata of varied quality and level of detail.
Since our focus is on SFX for notable visual events, we use the metadata to filter the SFX and video datasets to exclude categories outside the scope of our work. For audio, we removed human sounds (e.g., speech, crying) and music (e.g., saxophone, orchestra). For video, we aimed to keep visual scenes depicting sound-producing foreground objects and actions, removing largely ambient scenes (e.g., a video of an empty classroom is not of interest in our setup).
After filtering the PSE SFX library and segmenting the audio files, we obtained 332k SFX with a median duration of 3 seconds. We obtained 115k videos after similar filtering. We sample the videos at 1FPS, and extract up to 10 random frames per video, yielding approximately 1.1 million video frames to use in our automatic data curation pipeline.
We split the SFX into train, validation, and test splits, the latter two containing 4k SFX each with the rest used for training. The audio and visual data are both completely disjoint across the data splits.

\vspace{-0.1cm}
\subsection{Contrastive learning for SFX retrieval}
We use our HQ audio-visual training pairs to train a multimodal model that maps audio and video frames into a shared embedding space (Figure \ref{fig:block_diag}, right). Once trained, the model can be used for retrieval by computing the embedding distance between a query frame and a dataset of SFX to rank the SFX by their similarity to the visual query. Similar to recent work including Wav2CLIP \cite{wu2022wav2clip}, AudioCLIP \cite{guzhov2022audioclip}, and CLAP \cite{wu2023largescale}, we use contrastive learning with a standard contrastive loss \cite{oord2019representation} to train an audio encoder to project an audio signal into the CLIP image embedding space. We do so by keeping the original CLIP image encoder frozen during training, forcing the audio encoder to project audio into this pre-trained space. The choice of audio encoder is discussed in Section \ref{sec:model_design_choices}.

\vspace{-0.1cm}
\section{Experimental Design}
\label{sec:exp_design}
\subsection{Evaluation Datasets}
We evaluate our system on SFX retrieval from visual queries using three test datasets: two created using our automatic data curation pipeline (PSE-Test and ASFX-Test), and one ``in-the-wild'' dataset of YouTube videos (AVE-Test). 
\textbf{PSE-Test} is the test split described in Section \ref{sec:method}, containing 4k unique audio-visual pairs spanning 40 SFX categories. It is the most ``in-domain'' of the three test sets, though note again that it is completely disjoint from the training data in terms of the audio and video files used to create the pairs.
\textbf{ASFX-Test} was created similarly to PSE-Test, but using SFX from the publicly available Adobe Audition SFX Library\footnote{\url{https://www.adobe.com/products/audition/offers/AdobeAuditionDLCSFX.html.}} paired with the same HQ video dataset used above. It contains 5k unique audio-visual pairs across 19 SFX categories, and is completely disjoint from PSE-Test and the training data.
\textbf{AVE-Test} is a subset of the AVE dataset \cite{tian2018audio_AVE}, itself a curated subset of AudioSet \cite{gemmeke2017audioset} containing videos with a high degree of audio-visual correspondence. We filter AVE to exclude videos containing human sounds (notably speech) or music, which as noted earlier are outside the scope of this work, leaving 2.4k videos across 17 categories.\footnote{The list of SFX and videos used in evaluation is on our companion site.}
While our target application is matching HQ SFX to production-grade video, as represented by the first two datasets, AVE-Test allows us to evaluate our system's ability to generalize to ``in-the-wild'' noisy data.

\vspace{-0.1cm}
\subsection{Evaluation Metrics}
\label{sec:eval_metrics}
We use a given video frame as query to rank the audio from all test pairs based on their cosine similarity in the multimodal embedding space. We repeat this for all test pairs and compute standard retrieval metrics: Median Rank (MR) and Recall at 10 (R@10). A priori, we expect the metrics for matching the visual frame to the \textit{exact} audio file in the audio-visual pair to be low, since there could be many other audio files containing SFX that are also relevant to that frame. For this reason, we also compute category-level metrics using the categories from the SFX metadata, where a retrieved SFX is considered a correct match for a frame if its from the same SFX category as the exact audio from the audio-visual pair. Since R@10 is a more lenient metric at the category-level, we instead focus on Precision at 10 (P@10), i.e., the fraction of top-10 results that are from the right SFX category. This metric is most representative of our target use case, i.e., presenting the user with a short list of SFX, ideally \emph{all} of them from the right category given the visual query.

\vspace{-0.1cm}
\subsection{Baselines}
We evaluate our system against two strong pre-trained baselines:
Wav2CLIP \cite{wu2022wav2clip} and AudioCLIP \cite{guzhov2022audioclip}, both of which were trained to map audio input to the CLIP image embedding space, meaning they can be used for audio retrieval given a visual input in the same way as our system. 
\textbf{Wav2CLIP} uses a ResNet18 architecture for the audio encoder and was trained on 5-second chunks of video frames from VGGSound \cite{chen2020vggsound} sampled at a high frame rate and mean-pooled for segment-level CLIP embeddings. 
\textbf{AudioCLIP} uses an ESResNet \cite{guzhov2020esresnet} architecture initialized using ImageNet \cite{deng2009imagenet} pre-training for the audio encoder, and was trained on single random frames from AudioSet \cite{gemmeke2017audioset} videos. We use the AudioCLIP-Trimodal model in which the audio encoder was trained against CLIP image and CLIP text embeddings jointly. 
We use the original model weights provided by the authors for both baselines.

\begin{figure}[ht]
\centering
\includegraphics[width=\linewidth, trim={2cm 0 2cm 1cm}]{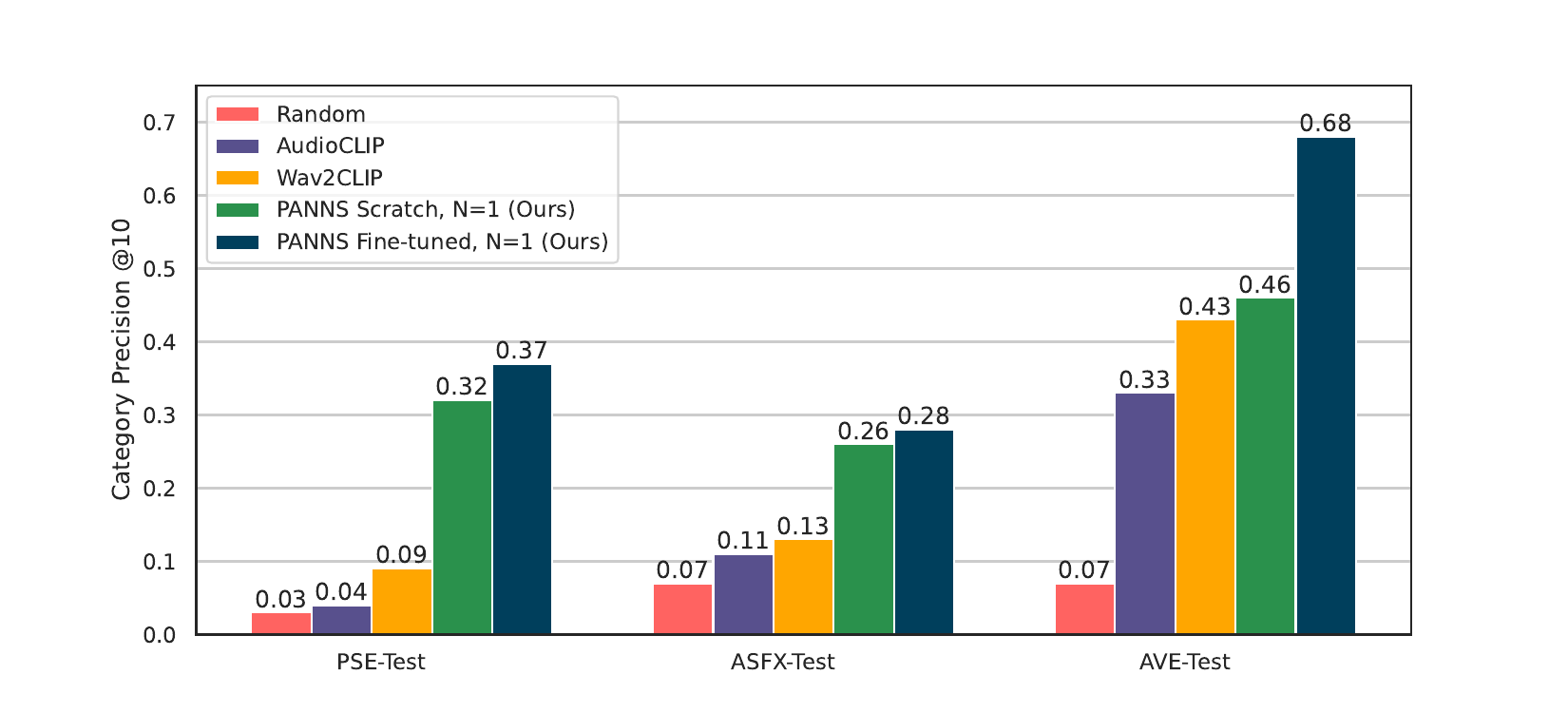}
\caption{SFX retrieval results: P@10 for our best model (from scratch and fine-tuned) and the baselines on the three test sets.}
\label{fig:overall_results}
\end{figure}

\begin{table*}[ht]
    \centering
    \footnotesize
    \bgroup
    \setlength{\tabcolsep}{5pt}
    \def\arraystretch{0.75}
    \begin{tabular}{lll|ll|ll|ll|ll|ll|ll}

         {} & {} & {} & \multicolumn{4}{c}{\textbf{PSE-Test}} & \multicolumn{4}{c}{\textbf{ASFX-Test}} & \multicolumn{4}{c}{\textbf{AVE-Test}} \\
         
         \cmidrule(lr){4-7}\cmidrule(lr){8-11} \cmidrule(lr){12-15}

         {}&{}&{} & \multicolumn{2}{c}{Exact} & \multicolumn{2}{c|}{Category} & \multicolumn{2}{c}{Exact} & \multicolumn{2}{c|}{Category} & \multicolumn{2}{c}{Exact} & \multicolumn{2}{c}{Category} \\

        {\textbf{Audio Encoder}} & {\textbf{Training}} & {\textbf{Pairing}} & MR & R@10 & R@10 & P@10 & MR & R@10 & R@10 & P@10 & MR & R@10 & R@10 & P@10\\
       
        \midrule
        
         PANNS \cite{kong2020panns} & Scratch & $\infty$ & 304 & 0.06 & 0.65 & 0.29 & 835 & 0.03 & 0.63 & 0.25 & 414 & 0.03 & 0.81 & 0.37 \\
         
         & Scratch  & 1 & 190 & 0.08 & 0.63 & 0.32 & 893 & 0.03 & 0.58 & 0.26 & 393 & 0.03 & 0.84 & 0.46\\
         
          & Fine-tuned & $\infty$ & 138 & 0.10 & 0.49 & 0.36 & \textbf{529} & 0.04 & 0.66 & \textbf{0.28} & 201 & \textbf{0.05} & 0.92 & 0.61\\

          & Fine-tuned  & 1 &  \textbf{122} & \textbf{0.11} & \textbf{0.66} & \textbf{0.37} & 597 & \textbf{0.03} & \textbf{0.61} & \textbf{0.28} & \textbf{193} & \textbf{0.05} & \textbf{0.95} & \textbf{0.68}\\

    \end{tabular}
    \egroup

 \caption{Ablation study over model design and data pipeline design choices.}
\label{tab:arch}

\end{table*}

\vspace{-0.1cm}
\subsection{Model Architecture and Training}
\label{sec:model_design_choices}
For our system, we experiment with two model architectures for the audio encoder: 
PANNS \cite{kong2020panns} and Wav2CLIP \cite{wu2022wav2clip}. 
AudioCLIP is not included due to its lower performance as a baseline compared to Wav2CLIP, as we discuss in Section~\ref{subsec:quant_and_qual}.
PANNS is a 14-layer CNN architecture with 81M trainable parameters. It was originally trained for audio classification on AudioSet. Wav2CLIP's ResNet18 audio encoder architecture has 11.2M trainable parameters, and was originally trained on VGGSound. For each of the two architectures, we try both training the model from scratch and fine-tuning the original model weights using our auto-curated training set. We train for a maximum of 150 epochs using batch size of 64 and learning rate of $10^{-5}$, and select the best epoch based on the Category Precision@10 metric on our validation split.

\vspace{-0.1cm}
\subsection{Automatic Data Curation Design Choices}
A key design choice in our automatic data curation pipeline is whether we allow the same video frame to be matched to multiple audio SFX (in the same data split), or whether we constrain the pairing. Since multiple SFX can end up with the same or similar natural language descriptions, if we pair them with the top matching video frame without any constraints, they may be paired to the same frame. This could lead to ``visual hubs'' that are potentially matched to hundreds of SFX. While this strategy ensures the best visual match for each SFX, it reduces the diversity of the dataset since we use fewer video frames overall. To address this, we can impose a matching constraint wherein each video frame can only be matched against $N$ SFX after which we remove it from the pool of available video frames. In the extreme case, we can set $N=1$, meaning each video frame can only be used in a single audio-visual pair, producing a dataset of completely unique audio-visual pairs. This has the effect of significantly increasing the diversity of the dataset, but at the cost of not always using the most closely matched video frame for each SFX. Having no constraint is equivalent to $N=\infty$, where a video frame can be matched to an unbounded number of SFX. In Section \ref{sec:res_data_ablation} we present an ablation study where we vary $N=\{1, 2, 5, 10, 100, \infty\}$ to generate variants of our training set and evaluate how they impact downstream retrieval performance.

\vspace{-0.1cm}
\subsection{User Study}
\label{sec:exp_design_user_study}
Since our goal is to produce a system the users find helpful, we complement our quantitative evaluation with a user study in which we compare our best performing model against the Wav2CLIP baseline. 
The user was presented with two videos, each containing the same single video frame but different SFX, one with the top SFX retrieved by our model and one with the top SFX from the baseline.
The user was asked to indicate ``Which video has a sound that better matches the image?' We deliberately did not play the full video from which the frame was taken, since precise synchronization of the SFX to the video is outside the scope of this work. Using a pool containing a random sample of 400 video frames from PSE-Test and 400 frames from AVE-Test, each user was shown a random selection of 30 frames from the pool as queries. The ordering of the two videos in each comparison was also randomized. We gathered 28 participants, and each completed 30 comparisons, yielding 840 total comparisons, evenly split between PSE-Test and AVE-Test.

\vspace{-0.1cm}
\section{Results}
\label{sec:results}

\vspace{-0.1cm}
\subsection{Quantitative and Qualitative Results}\label{subsec:quant_and_qual}
Based on the PSE \emph{validation} split, our best performing model uses the PANNS architecture for the audio encoder, starting from the pre-trained model weights and fine-tuning the encoder on our auto-curated training set which is generated with a pairing constraint of $N=1$ (for model selection we only tried $N=\{1, \infty\}$). In Figure~\ref{fig:overall_results} we present the P@10 scores for this model, the PANNS architecture trained from scratch, the two baselines and a random baseline on the three test sets described above. Our best model significantly outperforms the baselines across all three test sets. The results indicate two important takeaways: first, that our model is superior at matching HQ SFX to production quality video, our target application, as indicated by the results on PSE-Test and ASFX-Test. 
Second, and more surprising, our model successfully generalizes to noisy in-the-wild data, represented by AVE-Test, which is particularly interesting for our model trained from scratch as it was trained with \textit{only} HQ auto-curated audio-visual pairs. Conversely, the baselines, which were trained on large in-the-wild video collections, fail to generalize to the clean audio-visual data. This validates our proposed data curation pipeline as useful for training models that generalize to both clean HQ data and noisy in-the-wild data.

Turning to the qualitative results, the user study further validates our findings: participants preferred the SFX recommended by our model over the baseline 68.1\% and 66.4\% of the time for data from PSE-Test and AVE-Test, respectively.
Two binomial tests, one per dataset, indicated that these proportions were significantly higher than the expected 50\%, $p < 0.01$ (1-sided).

\vspace{-0.1cm}
\subsection{Model Ablation}
\label{sec:results_arch_ablation}
In Table \ref{tab:arch} we present an ablation over model design and data pipeline design choices. We conducted the same ablation using the Wav2CLIP audio encoder, but since results were consistently better for the PANNS encoder, we only present the latter. From initial experiments, we hypothesize that performance improves using PANNS vs.~Wav2CLIP due to both our HQ audio-visual data curation pipeline and increased model capacity (81M vs. 11.2M parameters). Fine-tuning a pre-trained PANNS works better than training from scratch, indicating we can benefit from transfer learning from AudioSet classification to our audio-visual matching task. Finally, when only comparing $N=1$ to $\infty$ for the pairing constraint in the automatic data curation pipeline, the former yields better results, especially for AVE, indicating it is preferable to have a more diverse training set, even at the cost of not pairing every SFX to its top-matching video frame. 
 
\vspace{-0.1cm}
\subsection{Automatic Data Curation Design Choice Ablations}
\label{sec:res_data_ablation}
To analyze the effect of the LLM in our data curation pipeline, we create a version of training data that uses the SFX metadata tags directly to retrieve visual frames via CLIP (as opposed to using natural language sentences generated from the tags using an LLM). We train our best performing architecture on this data and evaluate on the same test sets detailed previously. This version of the model performs similarly on PSE-Test and ASFX-Test but about 4pp worse on AVE-Test for Category Precision@10. This shows that using the LLM in our pipeline leads to overall better quantitative performance. Qualitatively, we observed that the usage of the LLM creates more realistic, action-focused audio-visual pairs than when tags are used directly.

To further explore the impact of the data pairing constraint on downstream retrieval performance, we train our best performing model design on six variants of the auto-curated training data. Each training set variant is generated with a different pairing constraint $N$ ranging from 1 to $\infty$ (no constraint). The results in Table \ref{tab:data_pair_ablation} show that for the scenario of matching HQ SFX to production-grade video, represented by PSE-Test and ASFX-Test, we can actually relax the pairing constraint, with optimal results for $N$ between 5-10, suggesting we can sacrifice some diversity in the training data for the sake of matching more SFX to their top ranked video frame. Conversely, if our goal is to generalize to in-the-wild noisy data, represented by AVE-Test, it is preferable to maintain a strict pairing constraint, indicating diversity is more important for generalizing to these data than maximizing pairing quality.

\begin{table}[tbh]
\centering
\footnotesize
\begin{tabular}{ccccccccc}
 \textbf{Pairing Limit} & \textbf{PSE-Val} & \textbf{PSE-Test} & \textbf{ASFX-Test} & \textbf{AVE-Test}   \\

\midrule

1 & 0.37 & 0.37 & 0.28 & \textbf{0.68} \\
2 & 0.39 & 0.38 & 0.29 & \textbf{ 0.68}  \\
5 & \textbf{0.40 }& \textbf{0.40} & 0.29 & 0.65  \\
10 & 0.39 & 0.39 & \textbf{0.30} & 0.64 \\
100 & 0.36 & 0.37 & \textbf{0.30} & 0.62 \\
$\infty$ & 0.35 & 0.36 & 0.28 & 0.61 \\

\end{tabular}

\caption{
Ablation over the pairing constraint in the automatic data-curation pipeline, using PANNS fine-tuned architecture. P@10.
}

\label{tab:data_pair_ablation}
\vspace{-1em}
\end{table}

\vspace{-0.2cm}
\section{Conclusion}
\label{sec:conclusion}
In this work we proposed a scalable automatic data curation pipeline for creating HQ audio-visual data, by leveraging powerful language and vision-language models. We showed that using the resulting data to train a contrastive multimodal model significantly outperforms baselines on the task of SFX retrieval given an input visual query, both quantitatively and qualitatively. Furthermore, we showed that our method generalizes not only to unseen data from the same setup, but also to ``in-the-wild'' noisy videos whereas the baselines, which are trained on noisy videos, fail to generalize to the HQ data. For future work we plan to extend our system to utilize video clips as queries instead of single video frames, which will enable us to leverage the motion cues in the video in addition to the semantic cues contained in a static frame for SFX retrieval.

\footnotesize
% \bibliographystyle{IEEEtran}
% \bibliography{refs23}

\end{sloppy}
\end{document}